\documentclass[aps,prl,amsfonts,amssymb,twocolumn,amsmath,preprintnumbers,nofootinbib,floatfix,superscriptaddress]{revtex4-1}%
\usepackage{siunitx}
\usepackage{graphicx}
\usepackage{mathrsfs}
\usepackage{bm}
\usepackage{amsmath}
\usepackage{dcolumn}
\usepackage{epstopdf}
\usepackage{dsfont}
\usepackage{amssymb}
\usepackage{tabularx}
\usepackage{array}
\usepackage{float}
\usepackage{color}
\usepackage{mathrsfs}
\usepackage{multirow}
\usepackage[T1]{fontenc}
\usepackage{makecell}
\usepackage[colorlinks,linkcolor=blue,anchorcolor=blue,citecolor=blue,urlcolor=blue]{hyperref}
\providecommand{\U}[1]{\protect\rule{.1in}{.1in}}

\newcommand{\vect}[1]{\boldsymbol{#1}}

\usepackage{amsfonts}
\usepackage{mathrsfs}
\usepackage{color}
\usepackage{graphicx}
\usepackage{bm}
\usepackage{amssymb}
\usepackage{xspace}
\usepackage{epstopdf}
\usepackage{dcolumn}
\usepackage{longtable}
\usepackage{multirow}
\usepackage{float}
\usepackage{comment}

\usepackage{pifont}
\newcommand{\cmark}{\ding{51}}
\newcommand{\xmark}{\ding{55}}

\makeatletter

\newcommand{\Rmnum}[1]{\expandafter\@slowromancap\romannumeral #1@}
\makeatother

\begin{document}

\title{Nonlinear Magnetoelectric Edelstein Effect}
\author{Jinxiong Jia}
\thanks{These authors contributed equally to this work.}
\affiliation{International Center for Quantum Design of Functional Materials, CAS Key Laboratory of Strongly-Coupled Quantum Matter Physics, and Department of Physics, University of Science and Technology of China, Hefei, Anhui 230026, China}
\affiliation{College of Physics and Optoelectronic Engineering, Shenzhen University, Shenzhen 518060, China}
\affiliation{Hefei National Laboratory, University of Science and Technology of China, Hefei 230088, China}
\author{Longjun Xiang}
\thanks{These authors contributed equally to this work.}
\affiliation{College of Physics and Optoelectronic Engineering, Shenzhen University, Shenzhen 518060, China}
\author{Zhenhua Qiao}
\email[]{qiao@ustc.edu.cn}
\affiliation{International Center for Quantum Design of Functional Materials, CAS Key Laboratory of Strongly-Coupled Quantum Matter Physics, and Department of Physics, University of Science and Technology of China, Hefei, Anhui 230026, China}
\affiliation{Hefei National Laboratory, University of Science and Technology of China, Hefei 230088, China}
\author{Jian Wang}
\email[]{jianwang@hku.hk}
\affiliation{International Center for Quantum Design of Functional Materials, CAS Key Laboratory of Strongly-Coupled Quantum Matter Physics, and Department of Physics, University of Science and Technology of China, Hefei, Anhui 230026, China}
\affiliation{College of Physics and Optoelectronic Engineering, Shenzhen University, Shenzhen 518060, China}
\affiliation{Quantum Science Center of Guangdong-Hongkong-Macao Greater Bay Area (Guangdong), Shenzhen 518045, China}
\affiliation{Department of Physics, The University of Hong Kong, Pokfulam Road, Hong Kong, China}

\begin{abstract}
    The linear Edelstein effect is a cornerstone phenomenon in spintronics that describes the generation of spin magnetization in response to an applied electric field.
    Recent theoretical advances have reignited interest in its nonlinear counterpart, the nonlinear Edelstein effect, in which spin magnetization is induced by a second-order electric field.
    However, the intrinsic contribution to both effects is generally forbidden in systems preserving time-reversal symmetry ($\mathcal{T}$) or composite symmetries such as $\mathcal{T}\tau_{1/2}$, where $\tau_{1/2}$ denotes a half-lattice translation.
    In such systems, spin magnetization typically emerges either from extrinsic mechanisms but limited to metals due to their Fermi-surface property, or from dynamical electric fields with a terahertz driving frequency.
    Here, we propose a new mechanism for spin magnetization, arising from the interplay of magnetic and electric fields, termed the nonlinear magnetoelectric Edelstein effect.
    Remarkably, its intrinsic component, determined purely by the material's band structure, can appear even in $\mathcal{T}$-invariant materials, but lacking inversion symmetry ($\mathcal{P}$), including insulators.
    On the other hand, we illustrate that its extrinsic component can serve as a sensitive indicator of the N\'eel vector reversal in $\mathcal{P}\mathcal{T}$-symmetric antiferromagnetic materials, offering a novel route for antiferromagnetic order detection.
    To validate our theory, we perform explicit calculations using a two-band Dirac model and a tight-binding model on a honeycomb lattice, finding that both effects yield sizable spin magnetization.
    Our findings establish the nonlinear magnetoelectric Edelstein effect as a versatile platform for both exploring nonlinear spin physics and enabling symmetry-based detection of antiferromagnetic order.
\end{abstract}

\maketitle

\noindent{\textit{\textcolor{blue}{Introduction.}}} ---
The Edelstein effect (EE) plays a pivotal role in spintronics, referring to the generation of nonequilibrium spin magnetization $\delta s$ in response to an applied electric field, characterized by the linear response coefficient $\alpha_{\alpha\beta}$ with $\delta s_\alpha=\alpha_{\alpha\beta}E_\beta$~\cite{Edelstein-solid_state_communication-1990,Aronov-1989,Kato-PRL-2004,silov-2004-APL,Cong_xiao-arXiv-2024}, where subscripts $\alpha, \beta$ are Cartesian indices and the Einstein summation convention is adopted.
In recent years, interest has expanded to its nonlinear counterpart, the nonlinear Edelstein effect (NLEE), in which spin magnetization is driven by a square electric field, $\delta s_\alpha=\alpha_{\alpha\beta\gamma}E_\beta E_\gamma$~\cite{Cong_Xiao-PRL-2023,Wang_Hua-PRB-2021,Cong_xiao-PRL-2022,Ju_Li-MTQ-2025,Amit-2024-arXiv,xiang-2025-arXiv}.
Crucially, since spin is odd under $\mathcal{T}$ symmetry, while the electric field is even, intrinsic contributions to both EE and NLEE are generally forbidden in $\mathcal{T}$-invariant systems and even in antiferromagnets with $\mathcal{T}\tau_{1/2}$ symmetry, where $\tau_{1/2}$ is the half lattice translation that does not operate on the physically observed quantities.
In such systems, spin magnetization typically arises either from an extrinsic mechanism~\cite{Cong_Xiao-PRL-2023,Ju_Li-MTQ-2025}, which relies on impurity scattering, or from the dynamical electric field~\cite{Cong_xiao-arXiv-2024}.
However, the former is a Fermi surface effect, which is inherently limited to metallic systems, while the latter requires terahertz-frequency electric fields, which are not always practical or feasible in experiments.
This raises a fundamental question: How does one obtain a DC-field-driven intrinsic spin magnetization in $\mathcal{T}$ or $\mathcal{T}\tau_{1/2}$-invariant systems, especially in insulating materials?

On the other hand, a central challenge in antiferromagnetic spintronics lies in detecting the orientation of the N\'eel vector, as conventional magnetic probes are ineffective due to the absence of net magnetization~\cite{Baltz-2018-RMP}.
To address this, recent studies have proposed using the nonlinear Hall effect~\cite{Yang_Gao-PRL-2014,Liang_Fu-PRL-2015,Binghai_Yan-PRL-2024,Di_Xiao-PRL-2021,Cong_xiao-PRL-2021,Liang_Fu-Hai_zhou_Lu-nature-2019,K_S_Burch-NM-2021,Suyang-Xv-science-2023,Hai_zhou_Lu-nature_review_physics-2021,Bernevig-arXiv-2024,Binghai_Yan-arXiv-2025} as an alternative detection method.
In antiferromagnets that break $\mathcal{P}\mathcal{T}$ symmetry, the extrinsic nonlinear Hall effect, characterized by the Berry curvature dipole (BCD) through the response relation $J_\alpha=\sigma_{\rm BCD}^{\alpha\beta\gamma}E_\beta E_\gamma$, has been shown to be sensitive to the N\'eel vector orientation~\cite{Tsymbal-PRL-2020}. 
However, since $\sigma_{\rm BCD}^{\alpha\beta\gamma}$ is a Fermi surface quantity, it vanishes in insulating systems.
In contrast, in $\mathcal{P}\mathcal{T}$-symmetric antiferromagnetic materials, the intrinsic nonlinear Hall effect, governed by the quantum metric dipole (QMD) through the relation $J_\alpha=\sigma_{\rm QMD}^{\alpha\beta\gamma}E_\beta E_\gamma$, has been proposed as a probe for detecting N\'eel vector reversal~\cite{Cong_xiao-PRL-2021,Di_Xiao-PRL-2021,Huang-PRL-2023}.
However, due to the permutation antisymmetry of $\sigma^{\alpha\beta\gamma}_{\rm QMD}$ in its first two indices, this effect is forbidden in many magnetic point groups, thereby limiting its general applicability.
These limitations motivate the search for a novel mechanism that can probe the orientation of the N\'eel vector in a broader class of antiferromagnetic materials.

In this Letter, we propose the nonlinear magnetoelectric Edelstein effect (NMEE) using Keldysh Green's function theory~\cite{Hua_Wang-PRB-2022,Book_nonequ_2013,Book2-2010}, response theory~\cite{Sipe0,Sipe1}, and semiclassical theory~\cite{Niu-PRB-1996, Niu-PRB-1999, Niu-RMP-2010, Yang_Gao-PRL-2014, Cong_xiao-PRL-2022, Cong_Xiao-PRL-2023, Jinxiong_Jia-PRB-2024, Xiangthird}.
The underlying mechanisms of NMEE can be classified into intrinsic and extrinsic types, depending on whether carrier scattering is involved or not.
We find that the NMEE arises from a novel quantum geometry tensor (QGT) in spin space, which is defined in analogy with the conventional QGT in momentum space.
Importantly, the intrinsic NMEE is symmetry-allowed in a broad class of non-centrosymmetric $\mathcal{T}$- or $\mathcal{T}\tau_{1/2}$-invariant materials, in sharp contrast to the intrinsic EE and NLEE, which typically require broken $\mathcal{T}$ symmetry.
Furthermore, since the intrinsic NMEE is a Fermi-sea effect, it remains nonzero in insulating antiferromagnets that lack $\mathcal{P}\mathcal{T}$ symmetry.
Meanwhile, the extrinsic NMEE can serve as a sensitive probe for the reversal of the N\'eel vector in antiferromagnetic materials in $\mathcal{P}\mathcal{T}$-symmetric systems, as extrinsic NMEE is $\mathcal{T}$-odd.
Compared to $\sigma_{\rm QMD}^{\alpha\beta\gamma}$, the extrinsic NMEE is allowed in a broader range of MPGs.
To validate our theory, we present explicit calculations using a two-band Dirac model and a tight-binding model on a honeycomb lattice.
Our work uncovers a new class of intrinsic nonlinear spin magnetization in $\mathcal{T}$-invariant systems and provides a novel mechanism to detect the orientation of the N\'eel vector in antiferromagnetic materials.

\bigskip
\noindent{\textit{\textcolor{blue}{Nonlinear magnetoelectric Edelstein effect}}} ---
We consider the spin magnetization induced by a uniform magnetic field $\vect{B}(t)=\vect{B}\sum_{\omega_\gamma=\pm \omega_B}e^{-i\omega_\gamma t}$ and electric field $\vect{E}(t)=\vect{E}\sum_{\omega_\beta=\pm \omega_E}e^{-i\omega_\beta t}$.
These fields enter the system via the Zeeman coupling $\hat{H}_B=\bar{\vect{B}}(t)\cdot \hat{\vect{\sigma}}$~\cite{Lorentz-force}, where $\bar{\vect{B}}=-g\mu_B\vect{B}$, with effective Land\'e factor $g$ and Bohr magneton $\mu_B$, and through the velocity gauge $\hat{H}_E=\vect{A}(t)\cdot \hat{\vect{v}}$, where the electric field is related to the vector potential via $\vect{E}(t)=-i\partial_t\vect{A}(t)$.
Following the standard Keldysh Green's function (GF) theory~\cite{Hua_Wang-PRB-2022,Book_nonequ_2013}, the $2\times 2$ matrix self-energy is diagonal, given by $\check{\Sigma}(t,t')=\hat{\Sigma}(t,t')\mathbb{I}_{2\times 2}$ with $\hat{\Sigma}(t,t')=\delta(t,t')(\hat{H}_E+\hat{H}_B)$.
The Dyson equation is given by $\check{\mathcal{G}}=\check{\mathcal{G}}_0+\check{\mathcal{G}}_0\star \check{\Sigma}\star\check{\mathcal{G}}$, where the $\star$ denotes the time convolution, $f\star g(t,t')=\int{\rm d}t_1f(t,t_1)g(t_1,t')$.
The $\check{\mathcal{G}}_0$ is zeroth order GF, $\check{\mathcal{G}}_0=[\hat{G}^r_0,\hat{G}^<_0;0,\hat{G}_0^a]$, where $\hat{G}_0^r$, $\hat{G}_0^a$ and $\hat{G}_0^<$ denote the retarded, advanced and lesser GFs, respectively.
By iterating the matrix Dyson equation twice, and using $\hat{G}_0^r(t,t')=[\hat{G}_0^a(t',t)]^\dag=-i\Theta(t-t')e^{-i\hat{H}_0(t-t')}$ and $\hat{G}_0(t,t')=if(\hat{H}_0)e^{-i\hat{H}_0(t-t')}$ with $f$ being the equilibrium Fermi-Dirac distribution, $H_0$ bare Hamiltonian and $\Theta$ the Heaviside step function, we obtain the matrix element of lesser GF up to the $EB$ order $\langle n\vect{k}|\hat{G}^{<,EB}_2|m\vect{k}\rangle=i\rho_{mn}^{ext}+i\rho_{mn}^{in}$, where $|n\vect{k}\rangle=e^{i\vect{k}\cdot \vect{r}}|u_{n\vect{k}}\rangle$ is the eigenstate of $\hat{H}_0$ with $|u_{n\vect{k}}\rangle$ being the periodic Bloch state, $\vect{k}$ crystal momentum, and $n$ band indices.
The $\rho_{mn}^{ext}$ is the extrinsic contribution and related to the first order of scattering time $\tau$, $\rho^{ext}_{mn}/E^\beta \bar{B}^\gamma=\tau \sum_\ell\left[ if_{m\ell}\sigma^\gamma_{m\ell}r^\beta_{\ell n}/\epsilon_{m\ell}+(m\leftrightarrow  n)^* \right]+i\tau f_{nm}\mathcal{D}_{mn}^\beta (\sigma^\gamma_{mn}/\epsilon_{nm})$, and  $\rho_{mn}^{in}$ is the intrinsic contribution and related to the zeroth order of scattering time $\tau$, $\rho^{in}_{mn}/E^\beta \bar{B}^\gamma=\sum_\ell[f_{n\ell}/\epsilon_{n\ell}-(m\leftrightarrow n)][\sigma^\gamma_{m\ell}r^\beta_{\ell n}+(m\leftrightarrow n)^*]/\epsilon_{nm}+if_{nm}/\epsilon_{nm}\mathcal{D}_{mn}^\beta(\sigma^\gamma_{mn}/\epsilon_{nm}) $.
Here, for simplicity, we only consider the DC limit of the electric and magnetic fields (the full lesser GF is provided in Supplemental Material~\cite{SM}). 
Here, $f_{nm}=f_n-f_m$, $\epsilon_{nm}=\epsilon_n-\epsilon_m$, $r_{nm}^\beta=\langle u_{nk}|i\partial_k^\beta|u_{mk}\rangle$ is the inter-band Berry connection, $\sigma_{nm}^\alpha=\langle u_{nk}|\hat{\sigma}^\alpha|u_{mk}\rangle$ and $\mathcal{D}_{mn}^\beta=\nabla_k^\beta-i(\mathcal{A}^\beta_m-\mathcal{A}_n^\beta)$ is the covariant derivative with $\mathcal{A}^\beta_n=\langle u_{nk}|i\partial_k^\beta|u_{nk}\rangle$ being the intra-band Berry connection.

Using the $\rho_{mn}^{ext}$ and $\rho_{mn}^{in}$, and definition of the spin magnetization, $\delta s^\alpha=\mu_B{\rm Tr}(-i\hat{\sigma}\hat{G}^{<,EB}_2)=\mu_B E_\beta \bar{B}_\gamma [\Gamma^{in}_{\alpha\gamma,\beta}+\tau\Gamma^{ext}_{\alpha\gamma,\beta}]$, we obtain the intrinsic NMEE $\Gamma_{\alpha\gamma,\beta}^{in}$ and extrinsic NMEE $\Gamma_{\alpha\gamma,\beta}^{ext}$ as:
\begin{align}
    \Gamma_{\alpha\gamma,\beta}^{in}&=eg\mu_B\sum_{n m}\int_{\vect{k}} \frac{4f_n}{\epsilon_{nm}^2}\left[\Delta\sigma^\gamma_{nm}\mathcal{Q}_{nm}^{\beta\alpha}+\frac{\Delta v_{nm}^\beta \mathcal{F}_{nm}^{\alpha\gamma}}{4\epsilon_{nm}}\right], \label{intrinsic_MNEE} \\
    \Gamma_{\alpha\gamma,\beta}^{ext}&=-\frac{eg\mu_B}{\hbar}\sum_{n m}^{m\ne n}\int_{\vect{k}}  f_n\nabla_\beta\frac{2\mathcal{S}_{nm}^{\alpha\gamma}}{\epsilon_{nm}}.\label{extrinsic_MNEE}
\end{align}
Here, $e$ and $\hbar$ are restored via dimension analysis, the three-band contributions in $\Gamma^{in}_{\alpha\gamma,\beta}$ are neglected that can be found in Supplemental Material~\cite{SM}, $\int_{\vect{k}}=\int_{\rm BZ} {\rm d}\vect{k}/(2\pi)^d$ where BZ stands for the Brillouin zone, $d$ is the spatial dimension, $\Delta \sigma_{nm}^\gamma=\sigma_{nn}^\gamma-\sigma_{mm}^\gamma$, $\Delta v_{nm}^\beta=v_n^\beta-v_m^\beta$, $v_n^\beta=\nabla_k^\beta \epsilon_{nk}$ and $\mathcal{Q}_{nm}^{\beta\alpha}={\rm Re}(r^\beta_{nm}\sigma^\alpha_{mn})$ is the Zeeman quantum metric~\cite{Longjun_Xiang-PRL-2025}.
In addition, $\mathcal{S}_{nm}^{\alpha\gamma}={\rm Re}(\Sigma_{nm}^{\alpha\gamma})$ and $\mathcal{F}_{nm}^{\alpha\gamma}=-2{\rm Im}(\Sigma_{nm}^{\alpha\beta})$ with $\Sigma_{nm}^{\alpha\beta}=\sigma^\alpha_{nm}\sigma^\beta_{mn}$.
Notably, although $\Sigma_{nm}^{\alpha\beta}$ arises from a nonlinear spin response, it can be naturally interpreted as a local QGT in spin space, as we demonstrate below.

Recall that the quantum distance between two Bloch states $|u_{n\vect{k}}\rangle$ and $|u_{m\vect{k}+d{\vect{k}}}\rangle=U_{d\vect{k}}|u_{m\vect{k}}\rangle$, is given by $ds^2_{nm}\equiv\left|\langle u_{n\vect{k}}|U_{d\vect{k}}|u_{m\vect{k}}\rangle\right|^2=g_{nm}^{\alpha\beta}dk_\alpha dk_\beta$ for $m\ne n$~\cite{Berry-1984,Bohm-2020quantum,provost1980riemannian,Longjun_Xiang-PRL-2025}, where $U_{d\vect{k}}=e^{-id\vect{k}\cdot\nabla_k}$ represents an infinitesimal momentum translation and $g_{nm}^{\alpha\beta}=r_{nm}^\alpha r_{mn}^\beta$ is the conventional QGT.
Inspired by this, we introduce an analogous quantum distance in spin space $ds^2_{nm}\equiv\left|\langle u_{n\vect{k}}|U_{d\vect{\theta}}|u_{m\vect{k}}\rangle\right|^2=\Sigma_{nm}^{\alpha\beta} d\theta_\alpha d\theta_\beta/4$, where $U_{d\vect{\theta}}=e^{-id\vect{\theta}\cdot\hat{\vect{\sigma}}/2}$ represents an infinitesimal spin rotation (detailed derivation is found in~\cite{SM}).
Here, $\Sigma_{nm}^{\alpha\beta}$ serves as the local QGT in spin space, which we term the $\mathcal{S}$-QGT.
In analogy with the conventional QGT, in which the real part $\mathcal{G}_{nm}^{\alpha\beta}={\rm Re}(g_{nm}^{\alpha\beta})$ and imaginary part $\Omega_{nm}^{\alpha\beta}=-2{\rm Im}(g_{nm}^{\alpha\beta})$ are denoted the quantum metric and Berry curvature, respectively.
Similarly, we define the $\mathcal{S}_{nm}^{\alpha\beta}$ and $\mathcal{F}_{nm}^{\alpha\beta}$ as $\mathcal{S}$-quantum metric and $\mathcal{S}$-Berry curvature, respectively.
Therefore, the extrinsic NMEE Eq.\eqref{extrinsic_MNEE} originates from the $\mathcal{S}$-quantum metric dipole $v_n\mathcal{S}_{nm}$ after integration by parts
and it clearly shows a Fermi surface property.
In contrast, the intrinsic NMEE Eq.\eqref{intrinsic_MNEE} stems from the Fermi-sea contribution, thus can persist even in insulating systems.

\bigskip
\noindent{\textit{\textcolor{blue}{Symmetry analysis.}}} --- 
To establish the symmetry criteria for the NMEE, we first examine the symmetry behavior of the $\mathcal{S}$-QGT.
Interestingly, we find that the $\mathcal{S}$-QGT shares similar symmetry behavior with the conventional QGT under $\mathcal{T}$, $\mathcal{P}$ and $\mathcal{P}\mathcal{T}$.
In particular, under time-reversal symmetry, both $\Omega_{nm}^{\alpha\beta}$ and $\mathcal{F}_{nm}^{\alpha\beta}$ are $\mathcal{T}$-odd, while $\mathcal{G}_{nm}^{\alpha\beta}$ and $\mathcal{S}_{nm}^{\alpha\beta}$ are $\mathcal{T}$-even, owing to $\mathcal{T}r^\alpha_{mn}(\vect{k})=r^\alpha_{nm}(-\vect{k})$ and $\mathcal{T}\sigma^a_{mn}(\vect{k})=-\sigma^a_{nm}(-\vect{k})$.
As a result, $\Gamma^{in}_{\alpha\gamma,\beta}$ is even under $\mathcal{T}$ and odd under $\mathcal{P}$, while $\Gamma^{ext}_{\alpha\gamma,\beta}$ is odd under both $\mathcal{T}$ and $\mathcal{P}$, as $\nabla_k$ is odd under $\mathcal{T}$ and $\mathcal{P}$.
To systematically derive the full constraints on the NMEE tensor from the magnetic point groups (MPGs), we invoke Neumann's principle~\cite{Neumann}, which requires that any physically observable tensor must remain invariant under the symmetry operations of the system.
This leads to the constraint:
\begin{align}
    \Gamma_{\alpha\gamma,\beta}^{in/ext}&=\eta_T \mathcal{R}_{\alpha\alpha'}\mathcal{R}_{\beta\beta'}\mathcal{R}_{\gamma\gamma'}\Gamma_{\alpha'\gamma',\beta'}^{in/ext}\label{eq:Neumann}
\end{align}
where $\mathcal{R}$ represents a point group operation and $\eta_T=+1 (-1)$ for $\mathcal{T}$-even (odd). 
The Tab.~\ref{tab1} summarizes the MPGs classified by the existence or absence of the intrinsic NMEE ($\Gamma^{in}_{\alpha\gamma,\beta}$) and extrinsic NMEE ($\Gamma^{ext}_{\alpha\gamma,\beta}$).
The intrinsic ($\sigma_{\rm QMD}^{\alpha\beta\gamma}$) and extrinsic ($\sigma_{\rm BCD}^{\alpha\beta\gamma}$) nonlinear Hall effect are also listed for comparison in later discussion~\cite{Bilbao}.

We wish to emphasize that NMEE is fundamentally distinct from the NLEE induced by the static electric field.
For example, the intrinsic NMEE is $\mathcal{T}$-even and $\mathcal{P}$-odd, while the intrinsic NLEE~\cite{Cong_xiao-PRL-2022} is $\mathcal{T}$-odd but $\mathcal{P}$-even.
Therefore, the intrinsic NMEE can occur in nonmagnetic materials, while the intrinsic  NLEE can only occur in systems without $\mathcal{T}$ symmetries.

\begin{table}[t!]
\caption{\label{tab1} Magnetic point groups classified by the existence or absence of intrinsic NMEE ($\Gamma^{ext}_{\alpha\gamma,\beta}$), extrinsic NMEE ($\Gamma^{in}_{\alpha\gamma,\beta}$), extrinsic nonlinear Hall effect $\sigma_{\rm BCD}^{\alpha\beta\gamma}$ and intrinsic nonlinear Hall effect ($\sigma_{\rm QMD}^{\alpha\beta\gamma}$).
}
\centering
\renewcommand{\arraystretch}{1.5}
\begin{tabularx}{\columnwidth}{>{\raggedright\arraybackslash}m{5.3cm}*4{>{\centering\arraybackslash}X}}
\hline
\hline
Magnetic point groups &$\sigma_{\rm BCD}$&$\sigma_{\rm QMD}$& $\Gamma^{ext}$&$\Gamma^{in}$\\
\hline
$6'/m,6'/mmm',m'\text{-}3',m'\text{-}3'm$&\xmark&\xmark&\cmark&\xmark\\
\makecell[l]{$-6,-6m2,-6m'2',23,4'32',-43m$}&\xmark&\xmark&\cmark &\cmark\\[5pt]
$6',6'22',6'mm'$&\cmark&\xmark&\cmark &\cmark\\[5pt]
\makecell[l]{$11',21',2221',231',m1',mm2',31',$\\$3m1',321',41',-41',4221',4mm1',$\\$-42m1',4321',-43m1',-61',61',$\\$6221',6mm1'$}&\cmark&\xmark&\xmark &\cmark\\[18pt]
$-6m21',432,-4'3m'$&\xmark&\xmark&\xmark &\cmark\\[5pt]
\makecell[l]{$-1',2/m',2'/m,-3',-3'm,-3'm',$\\$4/m',4'/m',4/m'mm,4'/m'm'm,$\\$4/m'm'm',6/m',6/m'mm,$\\$6/m'm'm'$}&\xmark&\cmark&\cmark &\xmark\\[17pt]
$m'mm,m'm'm',-6',-6'm2'-6'm'2$&\xmark&\cmark&\cmark &\cmark\\[6pt]
\makecell[l]{$1,2,2',222,2'2'2,m,m',mm2,$\\$m'm2',m'm'2,4,4',-4,-4',422,$\\$4'22',42'2',4m'm',-4'2m',4mm,$\\$4'm'm,-42m,-4'2'm,-42'm',6,$\\$6m'm',3,32,32',3m,3m'$}&\cmark&\cmark&\cmark &\cmark\\
\hline
\hline
\end{tabularx}
\end{table}

\bigskip
\noindent{\textit{\textcolor{blue}{Probing N\'eel vector via NMEE.}}} ---
The N\'eel vector and its reversed counterpart are related by time-reversal symmetry, making any $\mathcal{T}$-odd response inherently sensitive to the N\'eel vector reversal.
Recently, the intrinsic nonlinear Hall effect, $J_\alpha=\sigma_{\rm QMD}^{\alpha\beta\gamma}E_\beta E_\gamma$, has been proposed as a probe for detecting the N\'eel vector reversal in $\mathcal{P}\mathcal{T}$-symmetric antiferromagnets~\cite{Cong_xiao-PRL-2022,Di_Xiao-PRL-2021,Huang-PRL-2023}.
However, the $\sigma^{\alpha\beta\gamma}_{\rm QMD}$ is antisymmetric concerning the first two indices, which imposes stringent symmetry constraints and significantly limits the number of MPGs that allow a nonzero response.
In contrast, the extrinsic NMEE, $\Gamma^{ext}_{\alpha\gamma,\beta}$, is symmetric under exchange of the indices $\alpha$ and $\gamma$.
This key difference enables the extrinsic NMEE to exist in a broader set of MPGs, as shown in Tab.~\ref{tab1}.
Specifically, any MPGs that allow $\sigma_{\rm QMD}$ also permit the extrinsic NMEE.
Moreover, as shown in the first three lines of Tab.~\ref{tab1}, 13 MPGs support the extrinsic NMEE while strictly forbidding the $\sigma_{\rm QMD}$.
This distinction highlights the potential of extrinsic NMEE as a more versatile probe for detecting the N\'eel vector reversal in antiferromagnetic materials, where the intrinsic nonlinear Hall effect vanishes due to symmetry constraints.

In addition, Shao {\it et al.}~\cite{Tsymbal-PRL-2020} have demonstrated the extrinsic nonlinear Hall effect,  governed by the Berry curvature dipole is sensitive to the orientation of N\'eel vector in antiferromagnets lacking $\mathcal{P}\mathcal{T}$ symmetry, such as CuMnSb, which preserves the $\mathcal{T}\tau_{1/2}$ symmetry.
Since the symmetry properties of the intrinsic NMEE closely resemble those of the $\sigma_{\rm BCD}$ tensor, as summarized in Tab.~\ref{tab1}, it is reasonable to expect that the intrinsic NMEE would likewise be sensitive to the N\'eel vector orientation.
Furthermore, unlike $\sigma_{\rm BCD}$, which is a Fermi surface quantity and vanishes in insulating systems, the intrinsic NMEE can remain finite even in insulating antiferromagnetic materials.

\begin{figure}[t!]
\includegraphics[width=0.95\columnwidth]{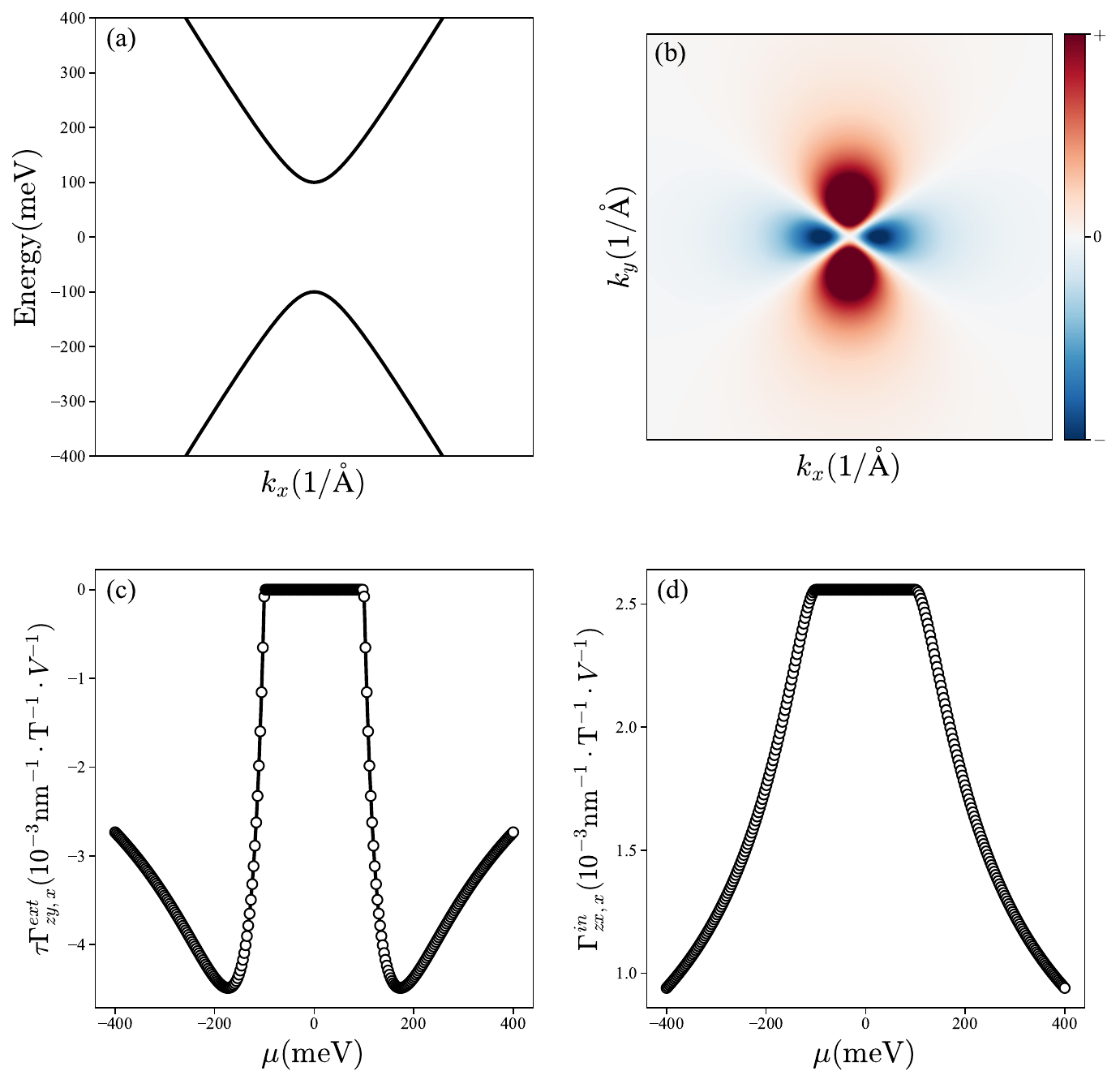}
\caption{(a) Band dispersions for Eq.~\eqref{model1}. (b) The $\vect{k}$-resolved distribution for integrand of $\Gamma^{in}_{zx,x}$ for the valence band. (c) The dependence of $\tau\Gamma^{ext}_{zy,x}$ on $\mu$. (d) The $\Gamma_{zx,x}^{in}$ as a function of $\mu$. Here, we choose the parameters $g$-factor $10$, $\tau=10{\rm fs}$, $\beta=0.1{\rm eV}$ and $v_F=0.6{\rm eV}\cdot$\AA.
} \label{FIG1}
\end{figure}

\bigskip
\noindent\textcolor{blue}{\textit{{Massive Dirac model.}}} ---
We first consider the massive Dirac model
\begin{align}
    H= v_F(k_y\sigma_x-k_x\sigma_y)+\beta\sigma_z,\label{model1}
\end{align}
where $k^2=k_x^2+k_y^2$, $\sigma_{x,y,z}$ are the Pauli matrices acting on the spin, $v_F$ is the Fermi velocity, and the last term is the mass term that breaks the time-reversal symmetry. 
The band dispersions of this model are $\epsilon_{\pm}=\pm d $ with $d=\sqrt{v_F^2k^2+\beta^2}$, as shown in Fig. \hyperref[FIG1]{1(a)}. 
Due to the broken $\mathcal{P}$ and $\mathcal{T}$ symmetries, both intrinsic and extrinsic NMEEs are allowed.

Since this model is the 2D system in $x-y$ plane, the NMEE tensor satisfies the constraint $\Gamma^{in/ext}_{\alpha\gamma,z}=0$, as $r_{mn}^z=0$.
Furthermore, the MPG of this system is $4m'm'$ with group generators $C_{4z}$ and $M_x\mathcal{T}$.
The $M_x\mathcal{T}$ operation imposes $\Gamma^{in}_{xxx}=\Gamma^{in}_{(xz,y)}=0$ and $\Gamma^{ext}_{(xx,y)}=\Gamma^{ext}_{(xx,z)}=0$, while the $C_{4z}$ rotation enforces $\Gamma^{in/ext}_{xx,x}=-\Gamma^{in/ext}_{yy,y}$, $\Gamma^{in/ext}_{(xx,y)}=\Gamma^{in/ext}_{(yy,x)}$, $\Gamma^{in/ext}_{(xz,x)}=\Gamma^{in/ext}_{(yz,y)}$ and $\Gamma^{in/ext}_{(xz,y)}=-\Gamma^{in/ext}_{(yz,x)}$, where $(xyz)$ denote a cyclic permutation of $xyz$.
Combining these symmetry constraints, we find that the intrinsic NMEE has two independent elements: $\Gamma^{in}_{zx,x}=\Gamma^{in}_{zy,y}$ and $\Gamma^{in}_{xz,x}=\Gamma^{in}_{yz,y}$ and the extrinsic NMEE has one independent element $\Gamma^{ext}_{zy,x}=-\Gamma^{ext}_{zx,y}$.
We focus on the in-plane magnetic field, under which the Lorentz force is negligible and $\Gamma^{in/ext}_{\alpha z,\beta}=0$.
Explicitly, we obtain $\mathcal{S}_{-+}^{zy}=\beta v_F k_x/d^2$, $\mathcal{F}_{-+}^{zx}=-2v_Fk_x/d$, $\sigma^x_{\pm}=\pm 2v_F k_y/d$, $\mathcal{Q}_{-+}^{xz}=-v_F^2k_y/(2d^2)$, substituting these QGTs into Eq.~\eqref{intrinsic_MNEE} and \eqref{extrinsic_MNEE}, we obtain the $\Gamma^{in}_{zx,x}$ and $\Gamma^{ext}_{zy,x}$ as: 
\begin{align}
    \Gamma^{in}_{zx,x}&=\frac{eg\mu_B}{6\pi v_F}\begin{cases}
        (3\mu^2-\beta^2)/(2|\mu|^3)&|\mu|>|\beta|\\
        1/|\beta|&|\mu|<|\beta|
    \end{cases}\label{model1.1}\\
    \Gamma^{ext}_{zy,x}&=\frac{eg\mu_B|\beta|}{4\pi\hbar v_F}\begin{cases}
        (\beta^2-\mu^2)/|\mu|^3&|\mu|>|\beta|\\
        0&|\mu|<|\beta|
    \end{cases}\label{model1.2}
\end{align}
As shown in Eq.~\eqref{model1.2}, the extrinsic NMEE vanishes when $\beta=0$, in which the system preserves the time-reversal symmetry, so that both intrinsic EE and intrinsic NLEE are forbidden, while the intrinsic NMEE remains allowed.
In Fig.~\hyperref[FIG1]{1(b)}, we show the $\vect{k}$-resolved distribution for the integrand of $\Gamma^{in}_{zx,x}$.
In Fig.~\hyperref[FIG1]{1(c)-(d)}, the dependence of $\tau\Gamma_{zy,x}^{ext}$ and $\Gamma_{zx,x}^{in}$ on the chemical potential are plotted, respectively.
In the gapped regime, we find that $\Gamma_{zy,x}^{ext}$ vanishes, while $\Gamma_{zx,x}^{in}$ remains non-zero. 
The latter is determined by the Fermi velocity $v_F$ and the band gap $\beta$ and shows the Fermi sea character.

\bigskip
\noindent\textcolor{blue}{\textit{{Tight binding model.}}} ---
We next apply our theory to a tight-binding model defined on a two-dimensional honeycomb lattice, which is given by~\cite{Qiao-PRB-2010}
\begin{align}
    H=&-t\sum_{\langle ij\rangle}c_{i\alpha}^\dag c_{j\alpha} +i\lambda_{R}\sum_{\langle ij\rangle}c_{i\alpha}^\dag(\vect{\sigma}_{\alpha\beta}\times \vect{d}_{ij})c_{j\beta}\notag\\
    &+\lambda\sum_{i}c_{i\alpha}^\dag\sigma^z_{\alpha\beta}c_{i\beta} \label{model2}
\end{align}
Here $c_{i\alpha}$ ($c_{i\alpha}^\dag$) is the annihilation (creation) operator for an electron with spin $\alpha$ at site $i$.
The first term is the nearest neighbor hopping. 
The second term corresponds to the nearest neighbor Rashba SOC which ensures the non-vanishing of $\mathcal{S}$-QGT.
Here, $\vect{d}_{ij}$ denotes a unit vector pointing from the site $j$ to $i$ with $\lambda_R$ representing the coupling strength, and $\vect{\sigma}$ is the vector of Pauli matrix acting on the spin space. 
The last term introduces a uniform exchange field.
The SOC term breaks the inversion symmetry and the nonzero exchange field breaks the time-reversal symmetry, so that both intrinsic and extrinsic NMEEs are allowed.
Furthermore, the MPG of this model is $6m'm'$, which is similar to the MPG $4m'm'$.
We identify one independent nonzero component for extrinsic NMEE: $\Gamma_{zy,x}^{ext}=-\Gamma_{zx,y}^{ext}$, and two independent components for intrinsic NMEE: $\Gamma^{in}_{zy,y}=\Gamma^{in}_{zx,x}$ and $\Gamma^{in}_{xz,x}=\Gamma^{in}_{yz,y}$.
As in the previous case, we only consider the in-plane magnetic field, so that $\Gamma^{in/ext}_{\alpha z,\beta}=0$.

\begin{figure}[t!]
\includegraphics[width=0.95\columnwidth]{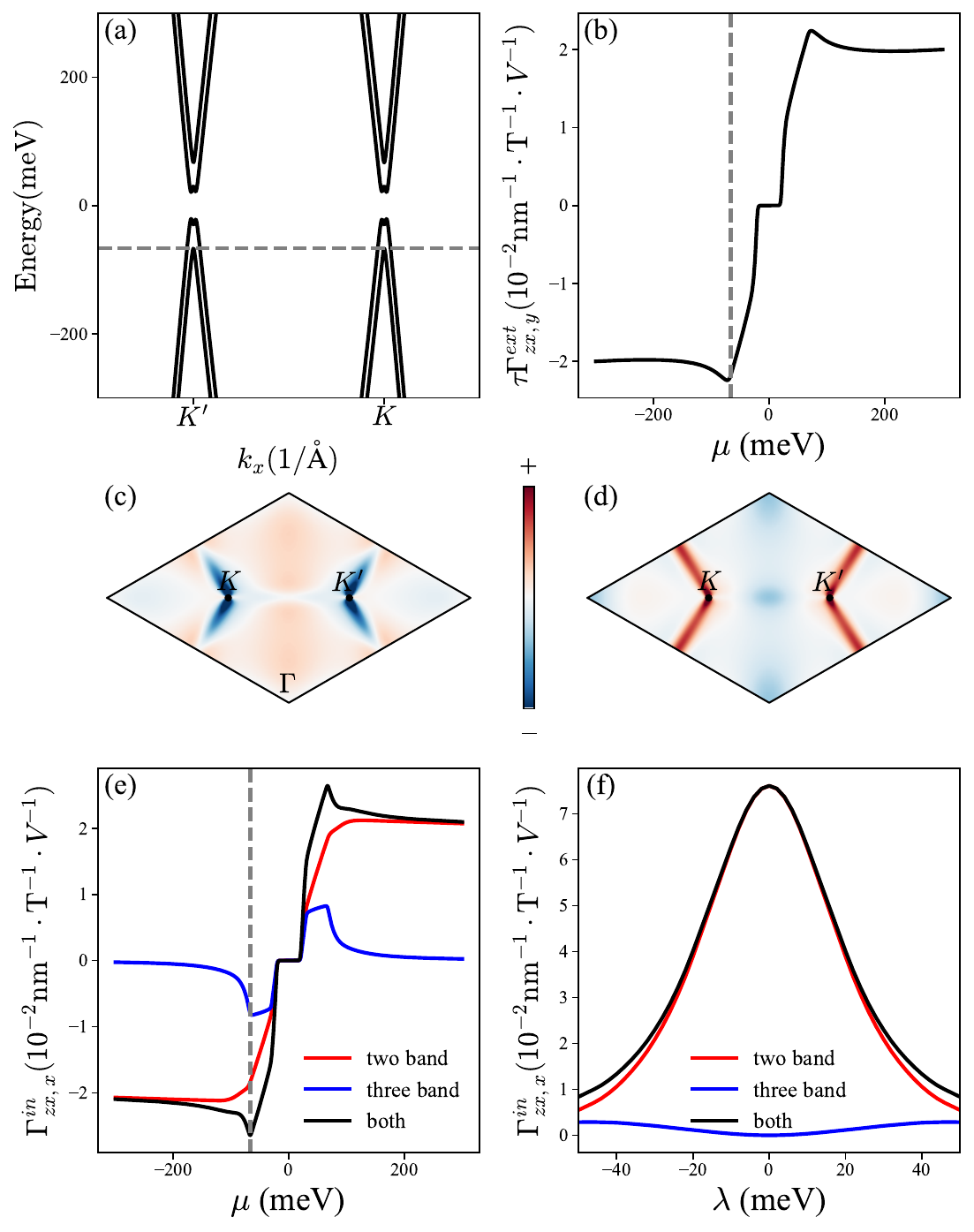}
\caption{(a) Energy band for this model. (b) The dependence of $\tau\Gamma^{ext}_{zx,y}$ on $\mu$ with $\tau=10{\rm fs}$. (c) The $\vect{k}$-resolved distribution for integrand of $\Gamma^{in}_{zx,x}$ with $n=1$. (d) The $\vect{k}$-resolved distribution for normalized $\mathcal{S}$-quantum metric dopole $\nabla_y\sum_m\mathcal{S}_{1m}^{zx}/\epsilon_{1m}$ (multiplied by a foctor of 10). (e) The dependence of intrinsic NMEE on $\mu$, where the red and blue lines denote the two-band and three-band contributions for Eq.~\eqref{intrinsic_MNEE}, respectively, and black line is the total intrinsic NMEE. (f) $\Gamma^{in}_{zx,x}$ versus $\lambda$ for $\mu=100{\rm meV}$. Here, we take $g=10$~\cite{Dai_Xi-g_factor}, $t=0.85{\rm eV}$, $\lambda_{R}=20{\rm meV}$, $\lambda=30{\rm meV}$.\label{FIG2}}
\end{figure}

In Fig.~\hyperref[FIG2]{2(a)}, the band dispersion for Eq.~\eqref{model2} is plotted, and the dashed line indicates the energy where the intrinsic and extrinsic NMEEs arrive at their maximum values. 
In Fig.~\hyperref[FIG2]{2(b)}, we calculate the $\tau\Gamma^{ext}_{zx,y}$ as the function of $\mu$ with $\tau=10{\rm fs}$~\cite{Cong_Xiao-PRL-2023}.
The Figs.~\hyperref[FIG2]{2(c)-(d)} show the distribution in the Brillouin zone for the integrand of $\Gamma^{in}_{zx,x}$ with $n=1$ and the normalized $\mathcal{S}$-quantum metric dipole $\nabla_y\sum_m\mathcal{S}_{1m}^{zx}/\epsilon_{1m}$, respectively.
One observes that these quantities are concentrated around the two inequivalent valleys $K$ and $K'$.
The Fig.~\hyperref[FIG2]{2(e)} plots the intrinsic NMEE, $\Gamma^{in}_{zx,x}$ as the function of $\mu$, and the two-band contribution is found be dominant.
Interestingly, although the intrinsic NMEE originates from the Fermi sea, it vanishes when the Fermi level lies within the band gap.
This behavior can be attributed to the presence of an additional combined symmetry $\mathcal{C}\mathcal{T}$ in the system, where $\mathcal{C}$ denotes the chiral symmetry, which is an anti-symmetry operation.
Hence, the $\mathcal{C}\mathcal{T}$ is an anti-unitary anti-symmetry operation, characterized by the transformation $U_{\mathcal{C}}U_{\mathcal{T}}H^*(\vect{k})U_{\mathcal{T}}^\dag U_{\mathcal{C}}^\dag=-H(-\vect{k})$, where $U_{\mathcal{T}}=-i\sigma_y$, $U_{\mathcal{C}}=\tau_z$ and $\tau_z$ is the Pauli matrix acting on the sublattice~\cite{SM}.
This combined symmetry ensures that the intrinsic NMEE contributions from the occupied and unoccupied states are equal, $\Gamma^{(in,{\rm occ})}_{\alpha\gamma,\beta}=\Gamma^{(in,{\rm unocc})}_{\alpha\gamma,\beta}$.
On the other hand, the total intrinsic NMEE, $\Gamma^{(in,{\rm occ})}_{\alpha\gamma,\beta}+\Gamma^{(in,{\rm unocc})}_{\alpha\gamma,\beta}$ is a fundamental quantity that always vanishes.
Combining these two facts immediately implies that $\Gamma^{(in,{\rm occ})}_{\alpha\gamma,\beta}=0$.
The Fig.~\hyperref[FIG2]{2(f)} show the dependence of $\Gamma^{in}_{zx,x}$ on the exchang field $\lambda$ for $\mu=100{\rm meV}$.
We observe that $\Gamma^{in}_{zx,x}$ arrived at the maximum when the system is time-reversal invariant ($\lambda=0$), and in which the three-band contribution can be neglected.

\bigskip
\noindent{\textit{\textcolor{blue}{Summary.}}} ---
In summary, we establish the quantum theory of the nonlinear magnetoelectric Edelstein effect and find that the NMEE arises from the $\mathcal{S}$-QGT defined in the spin space, which is similar to the conventional QGT in momentum space.
We further classify the NMEE into intrinsic and extrinsic types, depending on whether or not the mechanism is related to scattering time.
Importantly, the intrinsic NMEE is symmetry-allowed in $\mathcal{T}$, $\mathcal{T}\tau_{1/2}$-invariant systems, in contrast to the intrinsic EE and NLEE, which are generally forbidden in such systems.
Meanwhile, the extrinsic NMEE can serve as a more universal probe of the reversal of the N\'eel vector than the QMD in antiferromagnetic materials.
To facilitate the search for materials exhibiting the NMEE, a complete symmetry analysis is performed. 
Guided by this symmetry taxonomy, we evaluate the NMEE based on prototypical spin-orbit coupled model Hamiltonians.
From Fig.~\hyperref[FIG2]{2(b)}, one can see that applying scattering time $\tau=10{\rm fs}$, $0.1{\rm T}$ magnetic field and $10^5{\rm V/m}$ electric field yield a spin magnetization greater than $10^{-7}{\rm \mu_B\cdot nm^{-2}}$.
For intrinsic NMEE, a spin magnetization on the order of $10^{-7}\mu_B/{\rm nm}^2$ can be achieved by applying an electric field of $10^5{\rm V}/{\rm m}$ and a smaller magnetic field of $0.01{\rm T}$, as shown in Fig.~\hyperref[FIG2]{2(b)} and \hyperref[FIG2]{2(f)}.
These values are sizable, as existing techniques can detect spin magnetization down to $10^{-9}\mu_B/{\rm nm^3}$~\cite{Stern-PRL-2006,Kato-PRL-2004,Cong_xiao-arXiv-2024}.

\bigskip
\noindent{{\bf Acknowledgments}} ---
J.J. and Z.Q. are supported by the National Key R\&D Program of China (Grant No. 2024YFA1408103),
National Natural Science Foundation of China (No. 12474158, No. 12234017, and No. 12488101), 
Innovation Program for Quantum Science and Technology (2021ZD0302800),
Anhui Initiative in Quantum Information Technologies (AHY170000). 
J.W. and L.X. are supported by the National Natural Science Foundation of China (Grant No. 12034014, No. 12404059 and 12174262).
We also thank the Supercomputing Center of University of Science and Technology of China for providing the high-performance computing resources.


\begin{thebibliography}{00}
\bibitem{Lorentz-force}
Besides the spin degree of freedom, the magnetic field
is also introduced in the Hamiltonian by the minimal
coupling $\vect{A}\cdot\hat{\vect{v}}$.
However, by considering a two-dimensional
system pumped by an in-plane magnetic field,
the orbital effects can be neglected.
\end{thebibliography}

\end{document}